\documentclass[showpacs,aps,twocolumn]{revtex4}
\usepackage{graphicx,setspace}
\usepackage{bm}
\begin{document}
\title{Experimental study of nonlinear dust acoustic solitary waves in a dusty plasma}
\author{P. Bandyopadhyay*, G. Prasad, A. Sen and P. K. Kaw}
\address{Institute for Plasma Research, Bhat, Gandhinagar - 382428, India}
\date{\today}
\begin{abstract}
The excitation and propagation of finite amplitude low frequency solitary waves are investigated in an Argon plasma impregnated with kaolin dust particles. A nonlinear longitudinal dust acoustic solitary wave is excited by pulse modulating the discharge voltage with a negative potential. It is found that the velocity of the solitary wave increases and the width decreases with the increase of the modulating voltage, but the product of the solitary wave amplitude and the square of the width remains nearly constant. The experimental findings are compared with analytic soliton solutions of a model Kortweg-de Vries equation.
\end{abstract}
\maketitle
A dusty plasma, consisting of large macroparticles (usually the size of a few microns), immersed in an ionized gaseous medium with free floating electrons and ions, supports a rich variety of collective phenomena that have been the subject of a large number of investigations in recent times \cite{verheest_2001}. The massive dust particles, which also acquire a very high dynamic electric charge, bring about many significant changes in the overall collective behaviour  of the system including the creation of new modes. The dust acoustic wave (DAW) is one such new mode \cite{rao_1990} peculiar to this system and is an analog of the ion acoustic wave in a normal plasma with the dust particles playing the role of the ions by providing inertia and the pressure contributions for sustaining the wave coming from both the lighter species (the electrons and the ions). The linear characteristics of DAWs have by now been well established both theoretically \cite{mamun_2002} and experimentally \cite{merlino_97}. There is also an enormous theoretical literature on the topic of nonlinear DAWs (e.g. on the formation of solitons, vortices, shock structures etc.)\cite{rao_1998,volos_2002} but there has been relatively little experimental investigation \cite{nakamura_1999,samsonov_2002,samsonov_2004,pramanik} of these nonlinear entities in a dusty plasma. Nakamura \textit{et al.} \cite{nakamura_1999} showed that, an oscillatory ion-acoustic shock wave in a conventional two component Argon plasma transforms into a monotonic shock wave front when it propagates through a dusty plasma. Samsonov \textit{et al.} \cite{samsonov_2002} investigated the propagation and dissipation of longitudinal solitons in a 2D  monolayer hexagonal dust lattice of plastic monodisperse particles and invoked a linear chain theory, which included damping, dispersion and nonlinearity, to explain their experimental results. In a subsequent work the same group \cite{samsonov_2004} reported the experimental observations of shock melting in a two dimensional dusty plasma and their experimental results were supported by molecular dynamics simulations. Pramanik {\it et al} \cite{pramanik} observed dust acoustic turbulence arising from nonlinear mode-couplings and associated harmonics generation. While there have theoretical predictions of the existence of DAW solitons in many space plasma contexts, to the best of our knowledge there has been no laboratory studies of DAW solitary structures in a regular fluid state dusty plasma. In this Letter
we present the first experimental evidence of such long lived structures and describe their characteristics over a range of propagation velocities and spatial widths. We find that with the increase of the driving modulating voltage the solitary wave propagation velocity increases but its width decreases such that the product of the amplitude of the pulse and the square of its width remains nearly constant. We compare and interpret our results in terms of existing theoretical soliton solutions of a model Korteweg-deVries equation pertinent to our experimental parameter regime.\par
\begin{figure} 
\includegraphics[width=.45\textwidth]{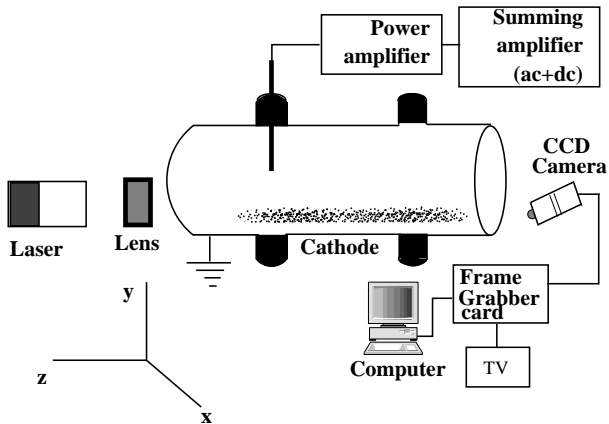}
\caption{A schematic of the experimental setup}
\label{fig:chamber}
\end{figure}

 
The experiments have been carried out in a set-up shown in Fig.~\ref{fig:chamber} which is similar to the set up described earlier in \cite{pintu_2007}. The stainless steel (SS) cylindrical chamber has eight radial ports and two axial ports for different purposes. The entire chamber is covered by a thin SS sheet to avoid stray arcing. Micron sized kaolin dust particles are sprinkled at the bottom of the SS cylindrical chamber. The vacuum chamber is initially pumped down to a base pressure of $10^{-3}$ mbar by a rotary pump. It is then purged with Argon gas using a precision needle valve and subsequently pumped down again to base pressure. This process is repeated several times. Finally the pressure is kept constant at 1 mbar. A discharge is then produced between a rod shaped anode and the grounded vessel (used as a cathode) at an applied voltage of $V_a=600$ volts. The applied voltage is then reduced to $542$ volts and the neutral gas pressure is gradually reduced to $0.09$ mbar to achieve the formation of a dense dust cloud. For such a discharge condition the average discharge current is usually $\sim 105 mA$. \par
The levitated dust particles are illuminated by a green $Nd-YAG$ diode laser. The laser light is spread into a sheet in the $x-z$ plane by a cylindrical lens (kept horizontally) and forward scattered light from the dust cloud is used to visualize the dust particles. The scattered light from the dust particles are captured using a CCD camera (25 fps) which is kept at an angle of $15^0$ to the $y-z$ plane. Further, the distance measurement in the horizontal plane ($x-z$) is duly corrected to account for the geometrical effects arising from the small but finite angle of inclination of the CCD camera with respect to the $z$ axis. Video frames are digitized by a frame grabber card with eight bit of intensity resolution and stored onto a high speed computer. The average dust temperature ($T_d$) is calculated from the random velocities of the particles by tracing single particle trajectories in different frames. The velocity, amplitude and width of the wave are all obtained from the still images. The plasma parameters like the ion density ($n_i$) and the electron temperature ($T_e$) are measured using a single Langmuir probe in the absence of dust particles inside a plasma. The electron density is estimated from the modified quasi-neutrality condition of the dusty plasma. A hot emissive probe is also used to measure the plasma potential and the floating potential. \par
\begin{figure}[h] 
\includegraphics[width=.45\textwidth]{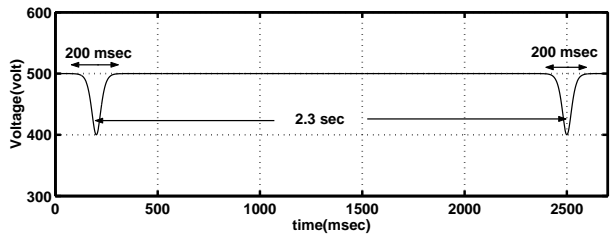}
\caption{\small{Typical voltage impulses used to excite dust acoustic solitary waves.}}
\label{fig:voltage}
\end{figure}
Once the dust cloud is formed a short negative voltage pulse is superimposed over the discharge voltage to perturb the equilibrium dust cloud. The dust cloud is then found to oscillate for the duration of the pulse length and to produce rarefaction and compressive structures corresponding to a dust acoustic wave. The oscillations quickly die down as soon as the pulse voltage stops. However for a voltage pulse that exceeds a certain threshold value (60 volts in this case) the dust column at the end of its oscillation phase launches a small density perturbation. This newly born compressive structure (a dust density perturbation) which constitutes a dust acoustic solitary wave (DASW) begins to move in the direction away from the anode and continues to move till the next pulse comes on (in this case about 1.5 to 2 secs later). The applied pulse voltage width and the time between successive pulses for the excitation of DASWs is adjusted such that the solitary wave can propagate over a long distance. The typical values of the pulse width and inter-pulse period are $200-300$ msec and $1.5-2.5$ sec respectively and are shown in Fig.~\ref{fig:voltage}. 

\begin{figure}[h] 
\includegraphics[width=.45\textwidth]{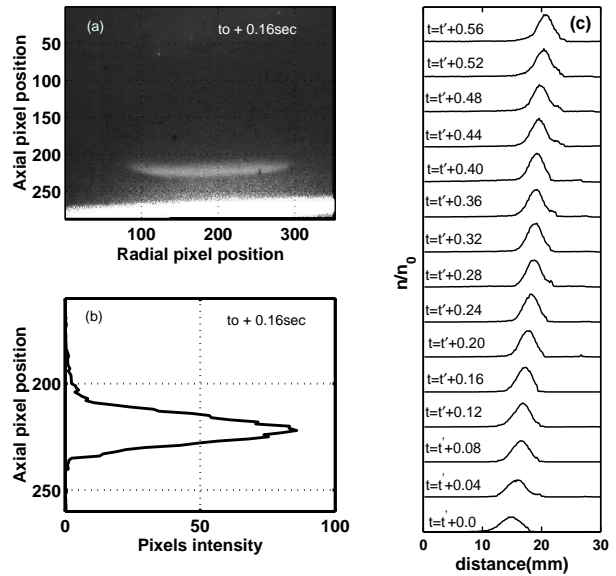}
\caption{\small{The propagation of dust acoustic solitary waves. a) video image of a typical solitary structure and b) its pixel intensity profile. c) Snapshots of the time evolution of the solitary wave at intervals of 0.04 sec.}}
\label{fig:typical}
\end{figure}
\begin{figure}[h] 
\includegraphics[width=.45\textwidth]{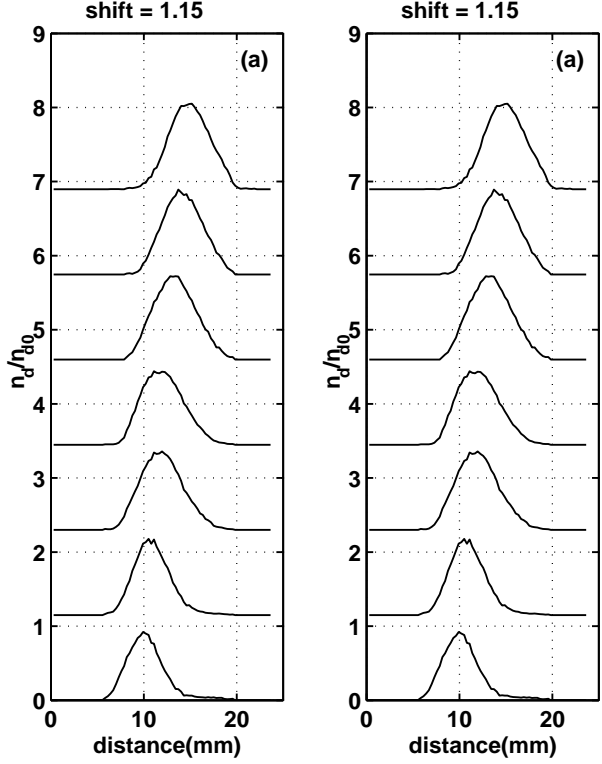}
\caption{\small{Solitary structures at two different excited pulse voltages for (a) $V_{pp}=60$ volts and (b) $V_{pp}=120$ volts. The wave propagates with higher velocity and amplitude but with lesser width when the applied voltage increases.}}
\label{fig:diff_amp}
\end{figure}
Figure \ref{fig:typical}(a), shows a typical solitary structure which propagates in the forward direction ($z$ direction). Figure \ref{fig:typical}(b) shows the variations of its pixel intensity (averaged over $x$) along  the $z$ direction. This pixel intensity profile is obtained after carrying out a background reduction and is proportional to the particle density. Figure \ref{fig:typical}(c), a typical time evolution plot of the solitary wave, shows that it  moves with a constant speed for a long distance at a constant pulse height. These time sequence plots are made at intervals of 0.04 sec. From this figure, it is also clear that the amplitude ($n_d/n_{d0}$) and the full width at half maximum (FWHM) of a typical DASW remains nearly constant throughout its journey. The $n_d$ and $n_{d0}$ profiles shown in Fig. \ref{fig:typical}(c) correspond to pixel intensities of DASWs and the equilibrium dust cloud respectively. The velocity of the DASWs is calculated from the distance traveled by a coherent structure between two consecutive frames.\par
\begin{figure}[hb] 
\includegraphics[width=.48\textwidth]{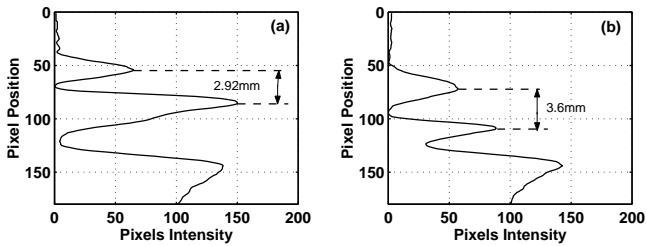}
\caption{\small{The breaking of large amplitudes structures into two different small amplitudes which moves with different velocities.}}
\label{fig:two_solitons}
\end{figure}
To further characterize the DASWs, we have varied the pulse height of the modulated discharge voltage by keeping the ON and OFF times of the pulse constant. The pulse height has been changed over a wide range from 60 to 120 volts. Figures \ref{fig:diff_amp}(a) and \ref{fig:diff_amp}(b) show typical DASW structures at two different peak voltages $V_{peak}$ = 60 and 120  volts respectively. In Fig.~\ref{fig:diff_amp}(a), a compressive structure moves 5 mm in 0.28 sec. In the other figure (Fig.~\ref{fig:diff_amp}(b)), the structure moves 10 mm in the same time interval. It is clear from these figures that the dust acoustic solitary waves move with different velocities if the excitation voltages are different. \par
We have also found that the amplitude of the compressive structure increases with the increase of the pulse height which is shown in Fig.~\ref{fig:diff_amp}(a) and Fig.~\ref{fig:diff_amp}(b). From these figures, it is also seen that the full width at half maximum (FWHM) decreases significantly with the increase of pulse height.\par 
In another set of experiments, we have observed the propagation of two solitary structures of different heights moving with different velocities. This happens when the driving pulse height exceeds a certain voltage (in this case it is 130 volts) and the resultant launched compressive pulse breaks up into two pulses. The amplitude of the leading pulse is seen to be a bit higher than the second one as shown in Fig.~\ref{fig:two_solitons}(a). With time the distance between these two solitary waves is seen to keep on increasing indicating that the higher amplitude pulse is traveling faster (see figures \ref{fig:two_solitons}(a) and \ref{fig:two_solitons}(b)). The above observations suggest that the dynamics of these solitary pulses are quite akin to theoretical results on compressive solitary waves of the Korteveg-de Vries kind.\par
To carry out a comparison of our experimental results with such theoretical solutions we consider the following model K-dV equation that has been derived in the context of dust acoustic waves \cite{rao_1998}.
\begin{eqnarray}  
&&\frac{\partial n_d ^{(1)}}{\partial \tau} + A n_d ^{(1)}\frac{\partial n_d ^{(1)}}{\partial \xi}+B\frac{\partial ^3 n_d ^{(1)}}{\partial \xi ^3}=0
\label{eqn:kdv}
\end{eqnarray}
where $n_d^{(1)}$ is the dust density perturbation normalized by the equilibrium dust density $n_{d0}$ and the coefficients are,
\begin{eqnarray}  
\nonumber
&&A=\frac{v_0^5}{(\delta-1)^2}\left[\delta^2+(3\delta+\sigma_i)\sigma_i+\frac{1}{2}\delta(1+\sigma_i^2)\right],\nonumber \\
\label{eqn:coeff}\\
&&B=\frac{v_0^3}{2}.
\nonumber
\end{eqnarray}
Here $\xi$ and $\tau$ are stretched coordinates such that $\xi=\epsilon^{1/2}(x-v_0t)$ and $\tau=\epsilon^{3/2}t$ with $\epsilon$ representing a smallness parameter measuring the weakness of the amplitude and $v_0$ is the wave phase speed normalized by the dust thermal velocity $C_{D}=(Z_dk_BT_i/m_d)^{1/2}$. The laboratory frame time ($t$) and space ($x$) variables are normalized by the dust plasma frequency $\omega_{pd}=(4\pi n_{d0}Z_d^2e^2/m_d)^{1/2}$ and the plasma Debye length $\lambda_d=(k_BT_i/4\pi n_0e^2)^{1/2}$. $Z_de=Q_d$ and $m_d$ are the dust charge and the mass of each dust grain respectively. Further  $\delta=n_{i0}/n_{e0}$ and $\sigma_i=T_i/Te$ where $n_{j0},T_{j}, (j=i,e)$ are the equilibrium densities and temperatures of the ions and electrons respectively.\par
A solitary pulse (soliton) solution of Eqn.( \ref{eqn:kdv}) results from a balance between the
nonlinear steepening induced by the convective term $n_d^{(1)}(\partial n_d^{(1)}/\partial \xi)$ and dispersive broadening due to the term $\partial ^3 n_d^{(1)}/\partial \xi ^3$. Such a stationary solution has the following exact analytic form, 
\begin{eqnarray}  
&&n_d^{(1)} = n_{dm}^{(1)}sech^2(\eta/\Delta_s)
\label{eqn:skdv}
\end{eqnarray}
where the amplitude $n_{dm}^{(1)} = 3\delta M/A $ and the width $\Delta_s = \sqrt{4B/\delta M}$ respectively. Here $\eta=\xi-\delta M\tau$, with $M=(v_0/v_{da}) = 1+\delta M$ being the Mach number with respect to the linear dust acoustic velocity $v_{da}$. 
To estimate the Mach number of the traveling pulse we have independently measured the linear phase velocity $v_{da}$ of the DAW by modulating the DC discharge voltage with a very low amplitude
sinusoidal voltage. Care was taken to ensure that the basic plasma conditions remained the same as those for which we have studied the solitary wave propagation. The velocity of this linear dust acoustic wave is then calculated from consecutive frames in the interval of 0.04 secs and turns out to be $\sim 1.67$ cm/sec for our discharge conditions. It is to be noted that all the constants of the expression $A$ (see eqn. (\ref{eqn:coeff})) are positive, making $A$ to be positive. It indicates that the dusty plasma supports dust acoustic solitary waves with a negative potential as $\phi_m^{(1)}=-v_0^2n_{dm}^{(1)}$.\par
\begin{figure}[hb] 
\includegraphics[width=.48\textwidth]{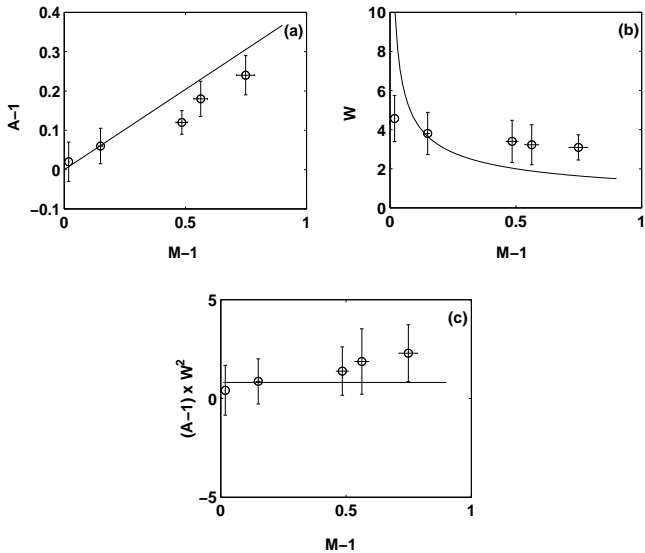}
\caption{\small{Variation of a) solitary amplitude, b) solitary width and c) product of solitary amplitude and the square of its width with Mach number. Solid line represents the solution of K-dV equation and open circles are the experimental points.}}
\label{fig:width}
\end{figure}
The width and amplitude of a K-dV soliton solution (\ref{eqn:skdv}) have a definite dependence on the pulse propagation speed. These theoretical dependencies (solid line) of amplitude, width and the product of the square of the width and the amplitude with the Mach number are plotted in Fig.~(\ref{fig:width})(a)--(c) along with experimental results (empty circles). All the constants were estimated using experimentally measured parameters like $Z_d=3\times10^3$, $T_e=8$ eV, $n_i=7\times10^{13}$ m$^{-3}$, $n_d=1\times10^{10}$ m$^{-3}$, $m_d=1\times10^{-13}$ kg. The ion temperature in our experiments is bit high (=0.3 eV) but consistent with similar earlier measurements reported  by Thompson \textit{et al.} \cite{thompson_1997}.  
As can be seen, the experimental results agree quite well with the theoretical estimates particularly for low amplitudes and low Mach numbers. The deviations at higher amplitudes (and Mach numbers) can be attributed to several factors. First of all the K-dV model itself ceases to be valid beyond a certain amplitude and corrections due to higher order nonlinearities
become important. Among other physical factors that could be contributing to the deviation one should comsider dust charge fluctuation effects as well as strong coupling effects, as have been pointed out in some past theoretical studies \cite{ghosh}. However a reliable estimate of these effects is difficult to quantify at this time because of experimental limitations (in making precise measurements of the involved quantities) as well as due to lack of reliable theoretical estimates of these effects in the nonlinear regime.\par 
To conclude, we report experimental observations of long lived low frequency solitary wave pulses in a laboratory dusty plasma. These pulses that correspond to nonlinear dust acoustic waves, are externally excited in a controlled manner and are investigated over a wide range of amplitudes and propagation velocities. Their dynamical properties are found to agree quite well, particularly at low amplitudes and low Mach numbers, with those of solitonic solutions of a model K-dV equation describing the propagation of nonlinear dust acoustic waves. The observed deviations at larger amplitudes indicate the influence of higher nonlinearities and/or the presence of additional physical effects that merit future experimental investigations. \par

\end{document}